\documentclass{bcp}

\usepackage{amsmath,amssymb,psfig,epsf}
\def\LaTeX{L\kern -.36em\raise .3ex\hbox{\sc a}\kern -.15em T\kern -.1667em%
\lower .7ex\hbox{E}\kern -.125em X}

\begin{document}

\mathclass{ }
\abbrevauthors{C. Sire and P.-H. Chavanis}
\abbrevtitle{Gravitational Collapse}

\title{Gravitational Collapse of a Brownian Gas}

\author{Cl\'ement Sire and Pierre-Henri Chavanis}
\address{Laboratoire de Physique Th\'eorique (UMR 5152 du CNRS), Universit\'e
Paul Sabatier,\\ 118, route de Narbonne, 31062 Toulouse Cedex 4, France\\
E-mail: {Clement.Sire{@}irsamc.ups-tlse.fr ~\&~
Chavanis{@}irsamc.ups-tlse.fr}}

\maketitlebcp

\abstract{We investigate a model describing the dynamics of a gas of
self-gravitating Brownian particles. This model can also have
applications for the chemotaxis of bacterial populations. We focus
here on the collapse phase obtained at sufficiently low
temperature/energy and on the post-collapse regime following the
singular time where the central density diverges. Several analytical
results are illustrated by numerical simulations.}

\section*{1. Introduction}
$$
$$
\vskip -0.5cm The general study of the static and dynamical properties
of a self-gravitating gas is a long standing problem in physics. Apart
for its clear applications in astrophysics, this problem has also many
conceptual interests, as for instance, the non-equivalence of
thermodynamical ensembles for many-body systems with long range
interactions \cite{paddy}.

In this paper, instead of considering the self-gravitating Newtonian
gas ({\it i.e.} obeying Newton's equations), we study a gas of
self-gravitating Brownian particles \cite{charosi} subject to a
friction originating from the presence of an inert gas and to a
stochastic force (modeling turbulent fluctuations,
collisions,...). This system has a rigorous canonical structure where
the temperature $T$ measures the strength of the stochastic
force. Thus, we can precisely check the thermodynamical predictions of
Kiessling \cite{kiessling} and Chavanis \cite{chavcano} obtained in
the canonical ensemble. 

For long-range interacting systems, the mean-field approximation is
presumed to become exact if the thermodynamical limit is taken
properly.  In a strong friction limit (or for large times), the
self-gravitating Brownian gas reduces to the Smoluchowski-Poisson
system, that is a Fokker-Planck equation describing the particle
density evolution in the presence of a gravitational field $\Phi$,
coupled self-consistently to Poisson's equation, stating that the
gravitational field is created by the gas itself.

This set of equations conserves mass and decreases the Boltzmann free
energy \cite{gt}.  They describe the competition between the
gravitational force which favors a collapsed state and the kinetic
pressure/diffusion/temperature which tends to spread the particles
over the entire accessible space. It is thus expected that below a
certain critical temperature $T_c$, the system undergoes a situation
of ``isothermal collapse'' \cite{chavcano}, which is the canonical
version of the ``gravothermal catastrophe'' \cite{lbw}.  These
equations have not been considered by astrophysicists because the
canonical ensemble is not the correct description of stellar systems
and usual astrophysical bodies do not experience a friction with a gas
(except dust particles in the solar nebula \cite{planete}).  Yet, it
is clear that the self-gravitating Brownian gas model is of
considerable conceptual interest in statistical mechanics to
understand the strange thermodynamics of systems with long-range
interactions and the inequivalence of statistical ensembles. In
addition, it provides one of the first model of stochastic particles
with long-range interactions, thereby extending the classical
Einstein-Smoluchowski model \cite{risken} to a more general context
\cite{gt}.

In addition, it turns out that the same type of equations occurs in
biology in relation with the chemotactic aggregation of bacterial
populations \cite{murray}. A general model of chemotactic aggregation
has been proposed by Keller \& Segel \cite{keller} in the form of two
coupled partial differential equations. In some approximation, this
model reduces to the Smoluchowski-Poisson system
\cite{charosi}.  Non-local
drift-diffusion equations analogous to the Smoluchowski-Poisson system
have also been introduced in two-dimensional hydrodynamics in relation
with the formation of large-scale vortices such as Jupiter's great red
spot \cite{rs,csr,houches}. These analogies, developed in
\cite{gt,analo}, give further physical interest to our Brownian model.

We now introduce a continuous mass density $\rho({\bf r})$ in a sphere a
radius $R$, and define respectively
\begin{itemize}
\item Total mass: $M=\int\rho({\bf r})\,d^{D}r$
\item Energy: $E=\frac{D}{2}MT+\frac{1}{2}\int \rho({\bf r})\Phi({\bf
r})\,d^Dr$
\item Entropy: $S=\frac{D}{2}M\ln T-\int \rho({\bf r})\ln[\rho({\bf
r})]\,d^Dr$
\end{itemize}
At equilibrium, the gravitational potential is given by the Boltzmann-Poisson
equation:
\begin{equation}
\Delta\Phi({\bf r})=S_D G \rho({\bf r}),\label{stat1}
\end{equation}
\begin{equation}
\rho({\bf r})={1\over Z}\exp[-\beta\Phi({\bf r})],\label{stat2}
\end{equation}
where $\Phi$ is the gravitational potential, $S_D$ is the surface of
the unit $D$-dimensional sphere, and $Z$ is the normalization
(partition function).  Note that these equations would be the same for
a Newtonian gas in the mean-field limit, as Eq.~(\ref{stat2}) simply
states that the density is given by the Boltzmann weight.
\begin{figure}
\centerline{ \psfig{figure=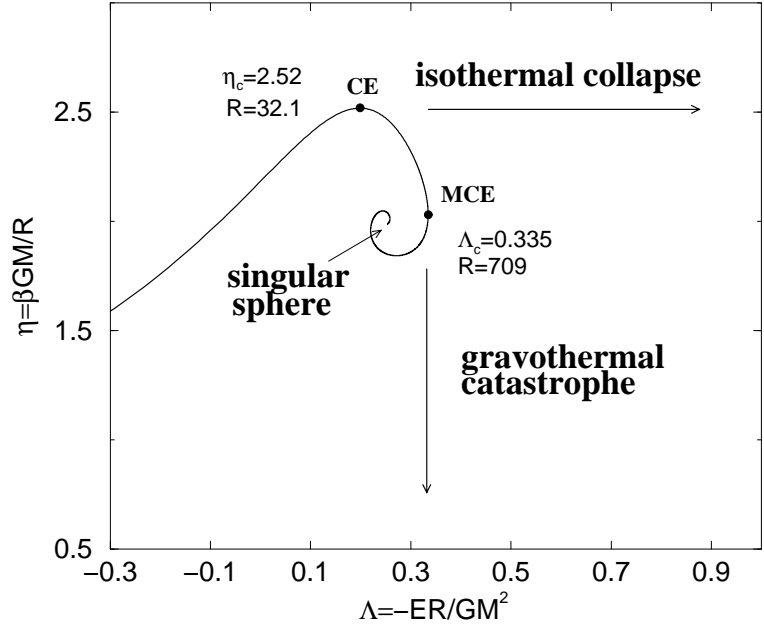,angle=0,height=8.5cm}}
\caption{We show the phase diagram of a self-gravitating Brownian
gas in $D=3$. In the canonical ensemble, there is no equilibrium
state below the rescaled temperature $T_c=\eta_c^{-1}\approx
0.396$, whereas in the microcanonical ensemble, the system
collapses below the rescaled energy $E_c=-\Lambda_c\approx
-0.335$. The equilibrium phase diagram in all dimensions $D$, as
well as for Langevin particles obeying Tsallis statistics has
been studied in \cite{cs1,cs2}.} \label{fig1}
\end{figure}
The microcanonical version of this model can be defined by imposing
the total energy $E$ instead of the temperature $T$. Although the
equilibrium distributions are the same, the stability limits differ in
microcanonical (fixed $E$) and canonical (fixed $T$) ensembles
\cite{chavcano}.  A typical phase diagram is displayed in
Fig.~\ref{fig1} (for $D=3$). This illustrates the fact that at low
enough temperature in the canonical ensemble, the system does not have
any equilibrium state, whereas the same situation arises in the
microcanonical ensemble at low enough energy. The static phase diagram
and the detailed stability analysis of the solutions have been
extensively studied in \cite{cs1,cs2} for isothermal and polytropic
distributions and for any dimension of space.

In this paper, we focus on the dynamical properties of the system when
no equilibrium state exists. In section 2, after stating the problem,
we study scaling collapse solutions in the canonical ensemble (for
$D\ge 2$, including the critical case $D=2$) and in the microcanonical
ensemble (for $D>2$). In both cases, the central density diverges in a
finite time $t_{coll}$ (except in $D=2$ for $T=T_{c}$ at which
$t_{coll}\rightarrow +\infty$). In section 3, we show that the
singular state reached at $t_{coll}$, which does not coincide with the
condensed state predicted by statistical mechanics
\cite{kiessling,chavcano}, is indeed not the final stage of the
dynamics. We then study the post-collapse stage, which is
characterized by the creation of a Dirac peak and the existence of a
backward scaling dynamics regime. Finally, we illustrate the very
large time regime where a dilute gas of Brownian particles evolves
around a massive core.

\section*{2. Collapse dynamics of self-gravitating Brownian particles}

\begin{itemize}
\item{\it Dynamics of the Smoluchowski-Poisson system}
\end{itemize}

At a given temperature $T$ controlling the diffusion coefficient, the
density $\rho({\bf r},t)$ of a system of self-gravitating Brownian
particles satisfies the following coupled equations:
\begin{eqnarray}
{\partial\rho\over\partial t}=\nabla\biggl \lbrack
{1\over\xi}({\cal D}\nabla\rho+\rho\nabla\Phi)\biggr\rbrack,
\label{brown1}
\end{eqnarray}
\begin{eqnarray}
\Delta\Phi=S_{D}G\rho,
\end{eqnarray}
where ${\cal D}$ is a diffusion constant. In the present paper, we
will consider ${\cal D}= T$ consistently with Einstein's relation. We
thus have to solve the Smoluchowski-Poisson system. At equilibrium,
implying a vanishing current, these equations simply reduce to
Eq.~(\ref{stat1}-\ref{stat2}).

\begin{figure}
\vskip 0.8cm
\centerline{ \psfig{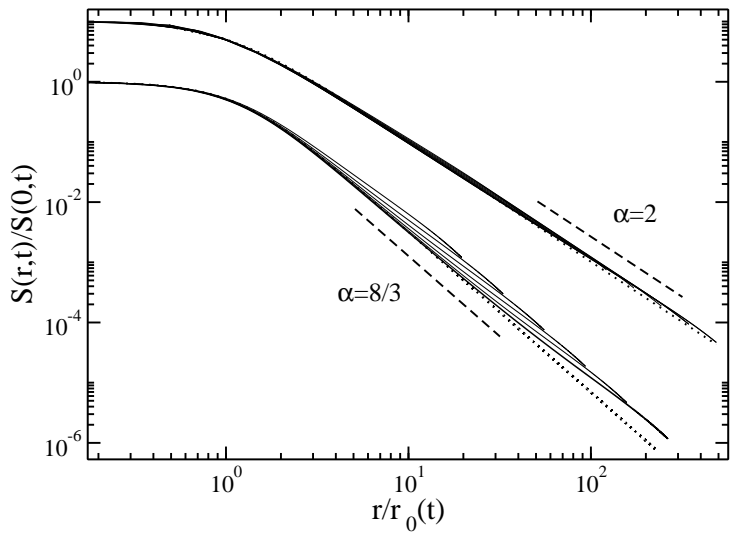}}
\caption{In $D=3$, we plot $s(r,t)/s(0,t)$ as a function of $r/r_0(t)$
for different times (density range $10^2-10^7$). The upper curve
corresponds to a constant diffusion coefficient ${\cal D}=T$,
i.e. $n=\infty$, leading to $\alpha=2$ \cite{charosi,cs1}. The lower
curve corresponds to ${\cal D}\sim \rho^{1/n}$ leading to
$\alpha=\frac{2n}{n-1}$ (the numerical simulations corresponds to
$n=4$, hence $\alpha=8/3$) \cite{cs2}. We compare the numerics to the
analytical scaling solutions (dashed lines).}
\label{fig2}
\end{figure}

The more general case where ${\cal D}$ is a function of $\rho$ itself
can also be of interest \cite{gt}. For example, by taking ${\cal
D}(\rho)\sim {\rho}^{1/n}$, one describes a gas of self-gravitating
Langevin particles displaying anomalous diffusion \cite{cs2}. The static
properties of this system reproduce that of a gravitational gas
described by Tsallis statistics.  The equilibrium distributions
correspond to polytropes which are associated with the $q$-entropy
$S_{q}=-{1\over q-1}\int (f^{q}-f) \,d^{D}{\bf r}\,d^{D}{\bf v}$,
where $f({\bf r},{\bf v})$ is the phase space density. The parameters
$q$ and $n$ are related to each other by the relation
\begin{equation}
n={D\over 2}+{1\over q-1}. \label{maxent10}
\end{equation}
This system can have self-confined equilibrium states depending
on $n$ and $D$. It can also undergo gravitational collapse at
sufficiently low temperature/energy as
illustrated on Fig.~\ref{fig2} (lower curve).

From now on, we take ${\cal D}=T$ and set $M=R=G=\xi=1$. We shall also
restrict ourselves to spherically symmetric solutions. The equations
of the problem then become
\begin{eqnarray}
{\partial\rho\over\partial t}=\nabla (T\nabla\rho+\rho\nabla\Phi),
\label{dim1}
\end{eqnarray}
\begin{eqnarray}
\Delta\Phi=S_{D}\rho,
\label{dim2}
\end{eqnarray}
with proper boundary conditions in order to impose a vanishing
particle flux on the surface of the confining sphere. These read
\begin{equation}
{\partial\Phi\over\partial r}(0,t)=0, \qquad \Phi(1)={1\over 2-D},
\qquad T{\partial \rho\over\partial r}(1)+\rho(1)=0,
\label{dim4}
\end{equation}
for $D>2$. For $D=2$, we take $\Phi(1)=0$ on the boundary. Integrating
Eq.~(\ref{dim2}) once, we can rewrite the Smoluchowski-Poisson system
in the form of a single integrodifferential equation
\begin{equation}
{\partial\rho\over\partial t}={1\over r^{D-1}}{\partial\over\partial
r}\biggl\lbrace r^{D-1}\biggl (T{\partial\rho\over\partial
r}+{\rho\over r^{D-1}}\int_{0}^{r}\rho(r')S_{D}r^{'D-1}dr'\biggr
)\biggr \rbrace.
\label{dim5a}
\end{equation}

The Smoluchowski-Poisson system is also equivalent to a single
differential equation
\begin{equation}
\frac{\partial M}{\partial t}=T \left(\frac{\partial^2 M}{\partial r^2}
-\frac{D-1}{r}\frac{\partial M}{\partial r}\right)
+{1\over r^{D-1}}M\frac{\partial M}{\partial r},
\label{sca}
\end{equation}
for the quantity
\begin{equation}
M(r,t)=\int_{0}^{r}\rho(r')S_{D}r^{'D-1}\,dr', \label{dint}
\end{equation}
which represents the mass contained within the sphere of radius
$r$. The appropriate boundary conditions are
\begin{equation}
 M(0,t)=N_0(t),\qquad M(1,t)=1,
\label{dintb}
\end{equation}
where $N_0(t)=0$, except if the density develops a condensed Dirac peak
contribution at $r=0$, of total mass $N_0(t)$.  It is also convenient to
introduce the function $s(r,t)=M(r,t)/r^{D}$ satisfying
\begin{equation}
{\partial s\over\partial t}=T\biggl ({\partial^{2}s\over\partial
r^{2}}+{D+1\over r}{\partial s\over\partial r}\biggr )+\biggl
(r{\partial s\over\partial r}+Ds\biggr )s.
\label{seq}
\end{equation}

\vskip 0.5cm
\begin{itemize}
\item{\it Self-similar solutions of the Smoluchowski-Poisson system in $D>2$}
\end{itemize}

In \cite{charosi,cs1,cs2}, we have shown that in the canonical
ensemble (fixed $T$), the system undergoes gravitational collapse
below a critical temperature $T_c$ depending on the dimension of
space.  The density develops a scaling profile, and the central
density grows and diverges at a finite time $t_{coll}$.

We look for self-similar solutions of the form
\begin{equation}
\rho(r,t)=\rho_{0}(t)f\biggl ({r\over r_{0}(t)}\biggr ), \qquad r_{0}=
\biggl ({T\over \rho_{0}}\biggr )^{1/2},
\label{dim5}
\end{equation}
where the King's radius $r_0$ defines the size of the dense core \cite{bt}.
In terms of the mass profile, we have
\begin{equation}
M(r,t)=M_{0}(t)g\biggl ({r\over r_{0}(t)}\biggr ), \qquad {\rm
with}\qquad M_{0}(t)=\rho_{0}r_{0}^{D},
\label{dim6}
\end{equation}
and
\begin{equation}
g(x)=S_{D}\int_{0}^{x}f(x')x^{'D-1}\,dx'. \label{dim7}
\end{equation}
In terms of the function $s$, we have
\begin{equation}
s(r,t)=\rho_{0}(t)S\biggl ({r\over r_{0}(t)}\biggr ), \qquad {\rm
with}\qquad S(x)={g(x)\over x^{D}}.
\label{dim6s}
\end{equation}

Substituting the {\it ansatz} (\ref{dim6s}) into Eq.~(\ref{seq}), we find that
\begin{equation}
{d\rho_{0}\over dt}S(x)-{\rho_{0}\over r_{0}}{dr_{0}\over dt}x
S'(x)={\rho_{0}^{2}}\biggl (S''(x)+{D+1\over
x}S'(x)+xS(x)S'(x)+DS(x)^{2}\biggr ),
\label{dim8}
\end{equation}
where we have set $x=r/r_{0}$. The variables of position and time
separate provided that $\rho_{0}^{-2}d\rho_{0}/dt$ is a constant
that we arbitrarily set equal to 2.  After time integration, this leads to
\begin{equation}
\rho_{0}(t)={1\over 2}(t_{coll}-t)^{-1},
\label{dim11}
\end{equation}
so that the central density becomes infinite in a finite time
$t_{coll}$, which appears as a integration constant.  The scaling
equation now reads
\begin{equation}
2S+xS'=S''+{D+1\over x}S'+S(xS'+DS).
\label{scalingd}
\end{equation}
For $D>2$, the scaling solution of Eq.~(\ref{scalingd}) was
obtained analytically in \cite{cs1} and reads
\begin{equation}
S(x)=\frac{4}{D-2+x^2},
\label{solscad}
\end{equation}
which decays with an exponent $\alpha=2$. This leads to (see
Fig.~\ref{fig2}, upper curve)
\begin{equation}
f(x)=\frac{4(D-2)}{S_D}\frac{x^2+D}{(D-2+x^2)^{2}} ,\qquad
g(x)=\frac{4x^D}{D-2+x^2}.
\end{equation}

Note finally that within the core radius $r_0$, the total mass  in
fact vanishes as $t\to t_{coll}$. Indeed, from  Eq.~(\ref{dim6}), we obtain
\begin{equation}
M(r_0(t),t)\sim \rho_{0}(t)r_{0}^{D}(t) \sim T^{D/2}(t_{coll}-t)^{D/2-1}.
\label{m0}
\end{equation}
Therefore, the collapse does {\it not} create a Dirac peak (``black hole'').

Near $T_c$, we find $t_{coll}\sim (T_c-T)^{-1/2}$, which is well
supported by numerical simulations \cite{charosi} and by a systematic
expansion procedure performed in \cite{tcoll}. In addition, the width of the
scaling regime is $\delta t\sim (T_c-T)^{1/2}$. This is an estimate of
the time $t_{coll}-\delta t$ from where the system enters the scaling
regime. Above $T_c$, the equilibration time characterizing the
exponential convergence to the stationary solution diverges like
$\tau\sim (T-T_c)^{-1/2}$ \cite{tcoll}.

In \cite{cs1}, we have also studied the collapse dynamics at $T=0$
for which we obtained
\begin{equation}
\rho_0(t)\sim S_D^{-1}(t_{coll}-t)^{-1},
\end{equation}
as previously, but the core radius is not given anymore by the
King's radius which vanishes for $T=0$. Instead, we find
\begin{equation}
r_0\sim \rho_0^{-1/\alpha},
\end{equation}
with
\begin{equation}
\alpha=\frac{2D}{D+2}.
\end{equation}
The scaling function $S(x)$ is only known implicitly
\begin{equation}
\left\lbrack\frac{2}{D+2}-S(x)\right\rbrack^{\frac{D}{D+2}}=K
x^{\frac{2D}{D+2}}S(x),
\label{st0}
\end{equation}
where $K$ is a known constant (see \cite{cs1} for details),
$S(0)=\frac{2}{D+2}$, and the large $x$ asymptotics $S(x)\sim f(x)\sim
x^{-\alpha}$. The mass within the core radius is now
\begin{equation}
M(r_0(t),t)\sim \rho_{0}(t)r_{0}^{D}(t) \sim (t_{coll}-t)^{D/2},
\label{m0t0}
\end{equation}
and it again tends to zero as $t\rightarrow t_{coll}$.  Comparing
Eq.~(\ref{m0}) and Eq.~(\ref{m0t0}) suggests that if the temperature
is very small, an apparent scaling regime corresponding to the $T=0$
case will hold up to a cross-over time $t_*$, with
\begin{equation}
t_{coll}-t_*\sim T^{D/2}.\label{tst0}
\end{equation}
Above $t_*$, the $T\ne 0$ scaling ultimately prevails.

\begin{itemize}
\item{\it Scaling and ``Pseudo-Scaling'' solutions  in $D=2$}
\end{itemize}

We now consider the critical dimension $D=2$ \cite{cs1}.  Above $T_c=1/4$, the
stationary solution can be explicitly computed, and the integrated
mass is found to be
\begin{equation}
M(r)=\frac{4T}{ 4T-1} \frac{r^2}{1+\frac{r^2}{4T-1}}.
\label{stat2d}
\end{equation}
Note that $M(1)=1$, ensuring that the whole mass is included in
this solution. Using $\rho=\pi^{-1}dM/d(r^2)$, we find that the
density profile is given by
\begin{equation}
\rho(r)=\frac{4\rho_0}{\pi}\frac{1}{(1+(r/r_0)^2)^{2}},
\label{rhost2d}
\end{equation}
with
\begin{equation}
 r_0=\sqrt{4T-1}\qquad {\rm and}\qquad \rho_0 r_0^{2}=T.
\label{r0stat}
\end{equation}
Note that the value of $T_{c}$ and the dependence of $r_{0}$ and
$\rho_{0}$ with the temperature coincide with the exact results
obtained within conformal field theory \cite{abdalla}. 

\begin{figure}
\centerline{ \psfig{figure=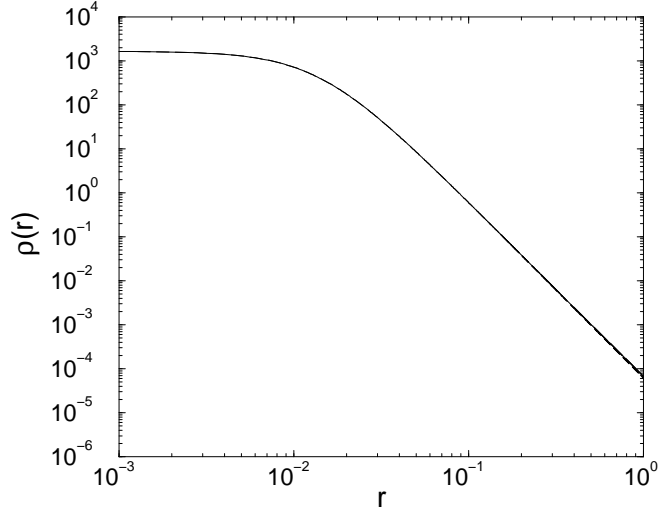,angle=0,height=7cm}}
\caption{At $T=T_c=1/4$, and when the central density has reached the
value $\rho(0,t)\approx 1644.8...=\frac{a(t)+1}{\pi}$ ($a(t)\approx
5166.3...$), we have plotted the result of the numerical calculation
compared to our exact scaling form
$\rho(r,t)=\frac{a(t)+1}{\pi}(1+a(t)r^2)^{-2}$ (dashed line). The two
curves are almost indistinguishable as the relative error is, as predicted,
of order $a^{-1}\sim 10^{-4}$. } \label{fig3}
\end{figure}

We now set the temperature to be exactly $T_c=1/4$. We then define
$a(t)$ in terms of the central density
\begin{equation}
\rho(0,t)=\frac{a(t)+1}{\pi}. \label{rhoa}
\end{equation}
The central density $\rho(0,t)$ is expected to diverge for
$t\rightarrow +\infty$, so that $a(t)$ is also expected to
diverge. Looking for a scaling solution for $M(r,t)$, we find that the
scaling function coincides with the expression of the stationary
solution of Eq.~(\ref{stat2d}) and Eq.~(\ref{rhost2d}). More
precisely, we find (see Fig.~\ref{fig3})
\begin{equation}
M(r,t)=\frac{a(t)r^2}{1+a(t)r^2}+a(t)^{-1}h(r,t),
\end{equation}
where $h(r,t)$ can also be computed perturbatively in $a^{-1}$
\cite{cs1}. The introduction of the next correction to scaling is
essential, as it governs the behavior of $a(t)$. Finally,
one obtains
\begin{equation}
\frac{da}{dt}=\frac{a}{\ln a-5/2}\left[1+O(\ln a^{-2})\right],
\label{dadt}
\end{equation}
leading to the asymptotic behavior
\begin{equation}
a(t)=\exp\left(\frac{5}{2}+\sqrt{2t}\right)\left[1+O(t^{-1/2}\ln
t)\right],
\label{at}
\end{equation}
and a similar behavior for the density according to Eq.~(\ref{rhoa}),
which is in perfect agreement with numerical simulations
\cite{cs1}. Note that at $T=T_c$ the central density does not diverge
in a finite time $t_{coll}$.

\begin{figure}
\centerline{ \psfig{figure=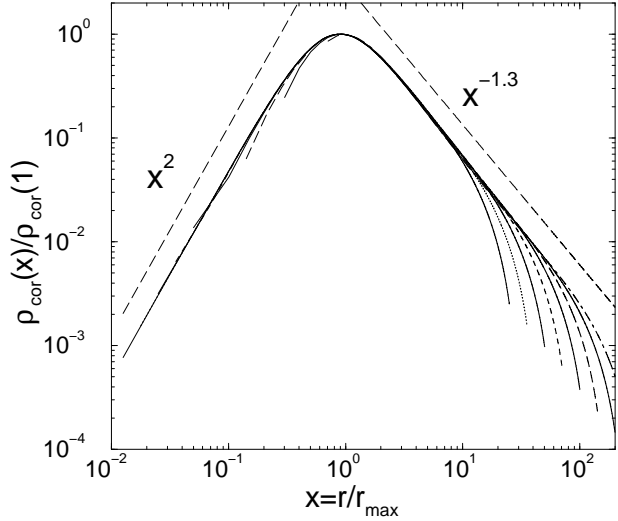,angle=0,height=7cm}}
\caption{At $T=T_{c}/2=1/8$, we have extracted the next correction
to scaling $\rho_{\rm cor}=\rho-T/T_c\,\rho_{T=T_c}$. We have then
plotted $\rho_{\rm cor}(r,t)/\rho_{\rm cor}(r_{\rm max}(t),t)$ as
a function of $x=r/r_{\rm max}(t)$, where $r_{\rm max}(t)$ is
defined as the location of the maximum of $\rho_{\rm cor}(r,t)$.
Consistently with the apparent scaling observed, we found $r_{\rm
max}^{-1}(t)\sim\sqrt{a}\sim\sqrt{\rho_{\rm cor}(r_{\rm
max}(t),t)}$. For $a=2^{n-1}{\times} 100$ ($n=1,...,8$), we have obtained
a convincing data collapse associated to $\alpha\approx 1.3$, in
agreement with the theoretical estimate for $\alpha$, in this
range of $a$. } \label{fig4}
\end{figure}

Strictly below $T_c$, the scaling equation of Eq.~(\ref{dim8})
does not have any global physical solution. However, we find that
by writing
\begin{equation}
M(r,t)=\frac{T}{T_c}\frac{a(t)r^2}{1+a(t)r^2}+M_{cor}(r,t),
\end{equation}
the correction $M_{cor}(r,t)$ adiabatically satisfies an effective
scaling relation of the form
\begin{equation}
M_{cor}(r,t)=a(t)^{\alpha/2-1}h_{cor}(\sqrt{a(t)}r),
\end{equation}
where $h_{cor}(x)\sim x^{2-\alpha}$, for large $x$. Hence, the
correction to the density scaling function satisfies
$\rho_{cor}(x)\sim x^{-\alpha}$. The index $\alpha$ is a very slowly varying
function of time, such that its explicit dependence on time does not
affect the scaling equation for $h_{cor}(x)$ or $\rho_{cor}(x)$
\cite{cs1}. Although the solution is not strictly speaking a true
scaling solution, the explicit dependence of $\alpha$ on $a(t)$ is so
weak that an apparent scaling should be seen with an effective
$\alpha$ almost constant for a wide range of values of $a(t)$ or
density. Hence, the total density profile is the sum of the scaling
profile obtained at $T_c$ with a $T/T_c$ weight (behaving as a Dirac
peak of weight $T/T_c$ at $t=t_{coll}$, as $\rho(r)\sim r^{-\alpha_c}$,
with $\alpha_c=4>D=2$) and of a pseudo-scaling solution associated to
an effective scaling exponent slowly converging to $\alpha=2$. More
explicitly, we find \cite{cs1}
\begin{equation}
2-\alpha(t)=2\sqrt{\frac{\ln\ln a}{2\ln a}} \left(1+O([\ln\ln
a]^{-1})\right). \label{expe}
\end{equation}
Let us illustrate quantitatively the time dependence of $\alpha$.  For
example, Eq.~(\ref{expe}) respectively leads to
$\alpha(a=10^3)=1.27...$, and to $\alpha(a=10^5)=1.34...$ Thus, for
the typical values of $a(t)$ accessible numerically of order $a\sim
10^5$, we expect to observe an apparent scaling solution with
$\alpha\approx 1.3$. This is confirmed by the scaling plot of
Fig.~\ref{fig4}.

Finally, for $T<T_c$, we note that the central density diverges again
in a finite time as one has
\begin{equation}
\frac{da}{dt}\sim a(t)^{1+\frac{\alpha(t)}{2}},
\end{equation}
implying 
\begin{equation}
\ln \rho_0(t)\sim -\ln(t_{coll}-t)\left[1+\sqrt{-\frac{\ln|\ln
(t_{coll}-t)|}{2\ln (t_{coll}-t)}}+...\right].
\end{equation}

\begin{itemize}
\item{\it Scaling solutions in the microcanonical ensemble $(D>2)$}
\end{itemize}
\begin{figure}
\vskip 0.7cm
\centerline{
\psfig{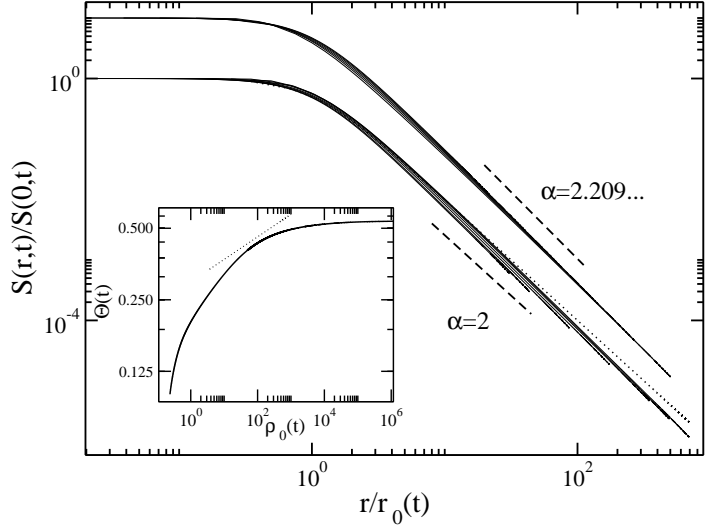}}
\caption{We plot $s(r,t)/s(0,t)$ as a function of $r/r_0(t)$ where
$r_0(t)\sim\rho_0(t)^{1/\alpha}$. We try both values $\alpha=2$ and
$\alpha=\alpha_{\rm max}=2.209733...$, and compare both data collapses
to the associated scaling function (dotted lines).  The scaling
associated to $\alpha_{\rm max}$ is clearly more convincing than that
for $\alpha=2$ (the two sets of curves have been shifted for
clarity). However, our simulations also suggest that $T(t)\sim
\rho_0(t)^{1-2/\alpha_{\rm max}}$ {\it does not diverge} at $t_{coll}$
(see the insert where a line of slope $1-2/\alpha_{\rm max}\approx
0.09491...$ has been drawn), so that the asymptotic scaling should
correspond to $\alpha=2$ (see also Guerra {\it et al.}
\cite{guerra}).}
\label{fig5}
\end{figure}

A microcanonical ensemble dynamics can be defined formally by
considering Eq.~(\ref{dim1}) with a uniform but time dependent
temperature $T(t)$ such that the total energy is kept strictly
constant. The resulting equations increase the Boltzmann entropy and
display a sort of ``gravothermal catastrophe'' \cite{charosi}.

If one looks for a scaling solution, one still has
$\rho_0(t)r_0^2(t)=T(t)$, but as the temperature is not necessarily
asymptotically constant near $t=t_{coll}$, the exponent $\alpha$
characterizing the decay of the density scaling function is not
determined by simple dimensional analysis. In Ref.~\cite{charosi}, we
found numerically that the scaling equation Eq.~(\ref{scalingd}) has
physical solutions only for $\alpha\le\alpha_{\rm max}$, with
$\alpha_{max}\simeq 2.209...$ for $D=3$. We also argued that the
system will select the exponent $\alpha_{max}$, since it leads to the
maximum increase of entropy.  This is illustrated on
Fig.~\ref{fig5}, where the scaling collapse is plotted assuming
respectively $\alpha=2$ and $\alpha=\alpha_{\rm max}$. Note that if
$\alpha>2$, the temperature diverges at $t=t_{coll}$, as $T(t)\sim
\rho_0(t)^{1-2/\alpha}$. However, in our most recent simulations (and
this was confirmed by Guerra {\it et al.} \cite{guerra}), we find that
as one approaches $t_{coll}$, the temperature ultimately saturates to
a finite value implying $\alpha=2$. This can be understood by observing 
that the conservation of energy implies that if the temperature
diverges, then the potential gravitational energy should also
diverge. However, if $\alpha<5/2$ (in $D=3$), it is straightforward
to show that the potential energy remains bounded. This exemplifies
one of the inherent flaws of a model for which the temperature is
maintained uniform, as we expect more realistically the temperature to
grow in the dense core (containing a vanishing mass according to 
Eq.~(\ref{m0})), while remaining finite in the halo. We have shown
\cite{cs2} that in a more realistic model where $T=T(r,t)$ is not necessarily
uniform, by a phenomenon similar to the one leading to solutions for
$\alpha\in [2;\alpha_{\rm max}]$, the dynamics now selects a unique
$\alpha>2$.

As the existence of solutions of  Eq.~(\ref{dim8}) for $\alpha>2$
is interesting and has physical implications in a more realistic
model with a non uniform temperature, let us mention some
analytical results about it. In the limit of large dimensions, the
scaling equation can be perturbatively solved (in powers of
$D^{-1}$).

Defining $z=\frac{DS(0)}{2}$ (which is of order $O(1)$), and
\begin{equation}
x_0^2=D+\frac{4}{z}+O(D^{-1}), {\rm \ or\ }
x_0=\sqrt{D}\left(1+\frac{2}{zD}+O(D^{-2})\right),
\label{x0d}
\end{equation}
such that $S(x_0)=\alpha/D$, and introducing
\begin{equation}
x_1^2=\frac{D}{z-1}+\frac{2(z-2)}{z(z-1)}+O(D^{-1}), \label{phix1}
\end{equation}
we have obtained the explicit form for the scaling function up to
second order in $D^{-1}$ \cite{cs1}
\begin{equation}
S(x)=\frac{\alpha}{D}\left[1+\left(1-\frac{\alpha}{2z}\right)
\left(\frac{x^2}{x_0^2}-1\right)
 \left(\frac{x^2}{x_1^2}+1\right)^{\frac{\alpha}{2}-1}\right]^{-1}.
\label{snextexp}
\end{equation}
The scaling exponent $\alpha$ is itself a function of $S(0)$ (or
$z$), defined by
\begin{eqnarray}
\alpha-2=\frac{4}{D}\left[\frac{1}{z}-\frac{2}{z^2}\right]
+\frac{8}{D^2}\left[\frac{5}{z}-\frac{26}{z^2}+
\frac{31}{z^3}-\frac{6}{z^4}\right.\nonumber\\
\left.-\left(\frac{1}{z}-\frac{7}{z^2}+
\frac{14}{z^3}-\frac{8}{z^4}\right)\ln z \right]+ O(D^{-3}).
\label{alphad2}
\end{eqnarray}
This function has a well defined maximum for
\begin{equation}
z=\frac{D}{2}S(0)=4+\left(\frac{41}{2}-6\ln 2\right)D^{-1}+ O(D^{-2}),
\label{zd2}
\end{equation}
associated to the value
\begin{equation}
\alpha_{\rm max}=2+\frac{1}{2}\,D^{-1}+\frac{11}{16}\,D^{-2}+O(D^{-3}).
\label{alphaexp2}
\end{equation}
This expansion gives $\alpha_{\rm max}=2.24...$ in $D=3$, in fair
agreement with the exact value $\alpha_{\rm max}=2.2097...$
obtained numerically in \cite{charosi}. In addition, the exponent
$\alpha=2$ is associated to $z=DS(0)/2=2+4D^{-1}+ O(D^{-2})$,
which coincides with the exact value $S(0)=4/(D-2)$ obtained in
Eq.~(\ref{solscad}).

\section*{3. Post-collapse dynamics in the canonical ensemble}

\begin{itemize}
\item{\it Post-collapse scaling at $T=0$}
\end{itemize}

According to general results of statistical mechanics
\cite{kiessling,chavcano}, the equilibrium state of self-gravitating
particles in the canonical ensemble is a Dirac peak containing all the
mass (for $D>2$). This is not the structure obtained at
$t_{coll}$. This implies that the evolution must continue in the
post-collapse regime. The scenario that we are now exploring
\cite{cs3} is the following. A central Dirac peak containing a mass
$N_0(t)$ emerges at $t>t_{coll}$, whereas the density for $r>0$
satisfies a scaling relation of the form
\begin{equation}
\rho(r,t)=\rho_{0}(t)f\biggl ({r\over r_{0}(t)}\biggr ),
\label{rhpost}
\end{equation}
where $\rho_0(t)$ is now decreasing with time (starting from
$\rho_0(t=t_{coll})\rightarrow +\infty$) and $r_0(t)$ grows with
time (starting from $r_0(t=t_{coll})=0$).  As time increases, the
residual mass for $r>0$ is progressively swallowed by the dense
core made of particles which have fallen on each other. It is the
purpose of the rest of this paper to show that this scenario
actually holds, as well as to obtain analytical and numerical
results illustrating this final collapse stage.

For $T=0$, the dynamical equation for the integrated mass $M(r,t)$
reads
\begin{equation}
\frac{\partial M}{\partial t}=
{1\over r^{(D-1)}}M\frac{\partial M}{\partial r},
\label{scaT0}
\end{equation}
with  boundary conditions
\begin{equation}
 M(0,t)=N_0(t),\qquad M(1,t)=1.
\end{equation}
We define  $\rho_0$ such that for small $r$
\begin{equation}
 M(r,t)-N_0(t)=\rho_0(t)\frac{r^D}{D}+...
\end{equation}
Up to the geometrical factor $S_D^{-1}$, $\rho_0(t)$ is the
central residual density (the residual density is defined as the
density after the central peak has been subtracted). For $r=0$,
Eq.~(\ref{scaT0}) leads to the evolution equation for $N_0$
\begin{equation}
\frac{d N_0}{d t}=\rho_0 N_0.
\label{N0T0}
\end{equation}
As $N_0(t)=0$ for $t\leq t_{coll}$, and since this equation is a
first order differential equation, it looks like $N_0(t)$ should
remain zero for $t> t_{coll}$ as well. However, since
$\rho_0(t_{coll})=+\infty$, there is mathematically speaking no
global solution for this equation and non zero values for $N_0(t)$
can emerge from Eq.~(\ref{N0T0}), as will soon become clear.

We then define
\begin{equation}
s(r,t)=\frac{M(r,t)-N_0(t)}{r^D},
\end{equation}
which satisfies
\begin{equation}
{\partial s\over\partial t}=\biggl
(r{\partial s\over\partial r}+Ds\biggr )s+\frac{N_0}{r^D}\biggl
(r{\partial s\over\partial r}+Ds-\rho_0\biggr ).
\label{sT0}
\end{equation}
By definition, we have also $s(0,t)=\rho_0(t)/D$.

We now look for a scaling solution of the form
\begin{equation}
s(r,t)=\rho_{0}(t)S\biggl ({r\over r_{0}(t)}\biggr ),
\label{st0post1}
\end{equation}
with $S(0)=D^{-1}$ and
\begin{equation}
\rho_0(t)= r_0(t)^{-\alpha},\label{r0t0}
\end{equation}
where $r_0$ is thus defined without ambiguity.  Inserting this scaling
{\it ansatz} in Eq. (\ref{sT0}), and defining the scaling variable
$x=r/r_0$, we find
\begin{equation}
{1\over \alpha\rho_0^{2}}\frac{d\rho_0}{dt}\left(\alpha S+x
S'\right)= S(DS+xS')+ \frac{N_0}{\rho_0r_0^D}{1\over x^{D}}(DS+xS'-1).
\label{scasT0}
\end{equation}
Imposing scaling, we find that both time dependent coefficients
appearing in Eq.~(\ref{scasT0}) should be in fact constant. We thus
define a constant $\mu$ such that
\begin{equation}
N_0=\mu \rho_0r_0^D.
\end{equation}
Equation (\ref{scasT0}) implies that $\rho_0\sim
(t-t_{coll})^{-1}$, which along with Eq.~(\ref{r0t0}) implies that
$N_0\sim (t-t_{coll})^{D/\alpha-1}$. We thus find a power law
behavior for $N_0$, which in order to be compatible with
Eq.~(\ref{N0T0}), leads to
\begin{equation}
\rho_0(t)=\left(\frac{D}{\alpha}-1\right)(t-t_{coll})^{-1}.
\end{equation}
We end up with the scaling equation
\begin{equation}
\frac{1}{D-\alpha}\left(\alpha S+x S'\right)+ S(DS+xS')+\mu
x^{-D}(DS+xS'-1)=0. \label{scas1T0}
\end{equation}
From Eq.~(\ref{scas1T0}), we find that the large $x$ asymptotics
of $S$ is $S(x)\sim x^{-\alpha}$. In a short finite time after
$t_{coll}$, it is clear that the large distance behavior of the
density profile ($r\gg r_0$) cannot dramatically change. We deduce
that the decay of $S$ should match the behavior for time slightly
less than $t_{coll}$ for which $S(x)\sim x^{-\frac{2D}{D+2}}$.
Hence the value of $\alpha$ should remain unchanged before and
after $t_{coll}$. Finally, we obtain the following exact behaviors
for short time after $t_{coll}$:
\begin{eqnarray}
\rho_0(t)&=&\frac{D}{2}(t-t_{coll})^{-1},\\
r_0(t)&=&\left(\frac{2}{D}\right)^{\frac{D+2}{2D}}
(t-t_{coll})^{\frac{D+2}{2D}},\\ N_0(t)&=&\mu
\left(\frac{2}{D}\right)^{\frac{D}{2}}
(t-t_{coll})^{\frac{D}{2}}.\label{n0postt0}
\end{eqnarray}
We note the remarkable result that the central residual density
$\rho(0,t)=S_D^{-1}\rho_0(t)$ displays a universal behavior just
after $t_{coll}$, a result already obtained in \cite{cs1}.
Moreover, we find that $N_0(t)$ has the same form as the mass
found within a sphere of radius $r_0(t)$ below $t_{coll}$, given
in Eq.~(\ref{m0t0}).

Moreover, the scaling function $S$ satisfies
\begin{equation}
\frac{D+2}{D^2}\left(\frac{2D}{D+2}S+x S'\right)+ S(DS+xS')+\mu
x^{-D}(DS+xS'-1)=0. \label{scas2T0}
\end{equation}
The constant $\mu$ is determined by imposing that the large $r$ behavior of
$s(r,t)$ (or $\rho(r,t)$) exactly matches (not simply
proportional) that obtained below $t_{coll}$, which depends on the
shape of the initial condition as shown in \cite{cs1}.
Equation (\ref{scas2T0}) can be solved implicitly by looking for
solutions of the form $x^D=z[S(x)]$. After cumbersome but
straightforward calculations, we obtain the implicit form
\begin{equation}
1+\frac{x^D}{\mu}S(x)=\left[1 +\frac{x^D}{\mu}\left(
S(x)+\frac{2}{D^2}\right)\right]^{\frac{D}{D+2}},
\end{equation}
which coincides with the implicit solution given in \cite{cs1}.  Note
that $S(x)$ is a function of $x^D$. We check that the above result
indeed leads to $S(0)=D^{-1}$, and to the large $x$ asymptotics
\begin{equation}
S(x)\sim
\mu^{\frac{2}{D+2}}\,\left(\frac{2}{D^2}\right)^{\frac{D}{D+2}}
x^{-\frac{2D}{D+2}}.
\end{equation}

Note finally that for $T=0$, $N_0(t)$ saturates to 1 in a finite
time $t_{end}$, corresponding to the deterministic collapse of the outer
mass shell initially at $r=1$. Indeed, using the Gauss theorem, the
position of a particle initially at $r(t=0)=1$ satisfies
\begin{equation}
\frac{dr}{dt}=-r^{-(D-1)}.
\end{equation}
The position of the outer shell is then
\begin{equation}
r(t)=(1-Dt)^{1/D},
\end{equation}
which vanishes for $t_{end}=D^{-1}$.

\vskip 0.5cm
\begin{itemize}
\item{\it Post-collapse scaling at $T>0$}
\end{itemize}

\begin{figure}
\vskip 0.7cm
\centerline{
\psfig{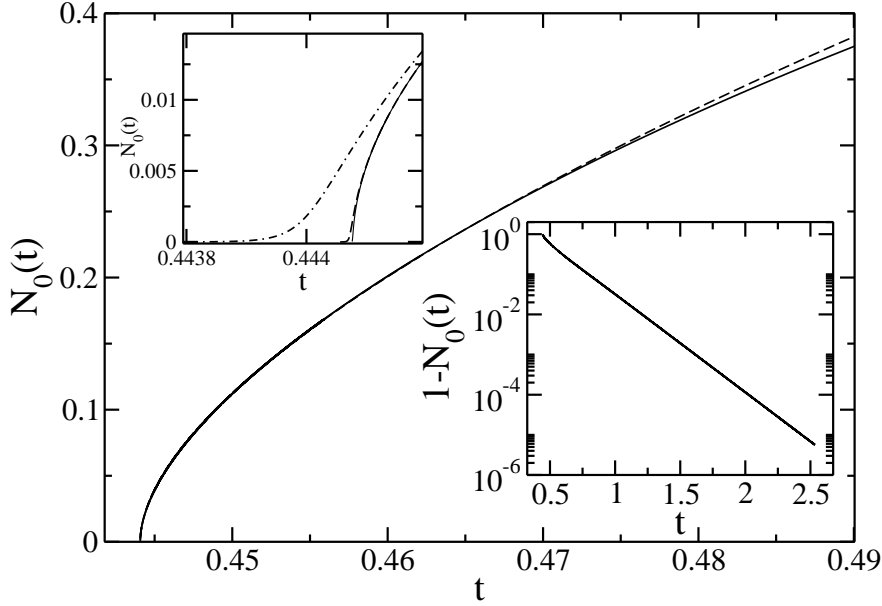}}
\vskip 0.5cm
\caption{We plot $N_0(t)$ for small time (full line). This is compared
to $N_0(t)^{\rm Theory}{\times}\left[1+a(t-t_{coll})^b\right]$ (dashed
line), where $N_0(t)^{\rm Theory}$ is given by Eq.~(\ref{n0post}) with
$\mu=8.38917147...$, and $a\approx 1.7.$ and $b\approx 0.33$ are
fitting parameters. Note that the validity range of this fit goes well
beyond the estimated $t_*$ with $t_*-t_{coll}\sim T^{D/2}\sim
0.09$. The bottom insert illustrates the exponential decay of
$1-N_0(t)\sim{\rm e}^{-\lambda t}$. The best fit for $\lambda$ leads
to $\lambda\approx 5.6362$ to be compared to the eigenvalue computed
by means of Eq.~(\ref{eigen1}), $\lambda= 5.6361253...$. Finally, the
top inset illustrates the sensitivity of $N_0(t)$ to the space
discretization $dx$, which introduces an effective cut-off necessary
in order to smoothly cross the singularity at $t=t_{coll}$ (a factor 4
in $dx$ between each of the 3 curves). Note the small time scale~: even
the curve corresponding to the coarsest discretization becomes
indistinguishable from the others for $t>0.448$.}
\label{fig6}
\end{figure}

\begin{figure}
\vskip 0.7cm
\centerline{
\psfig{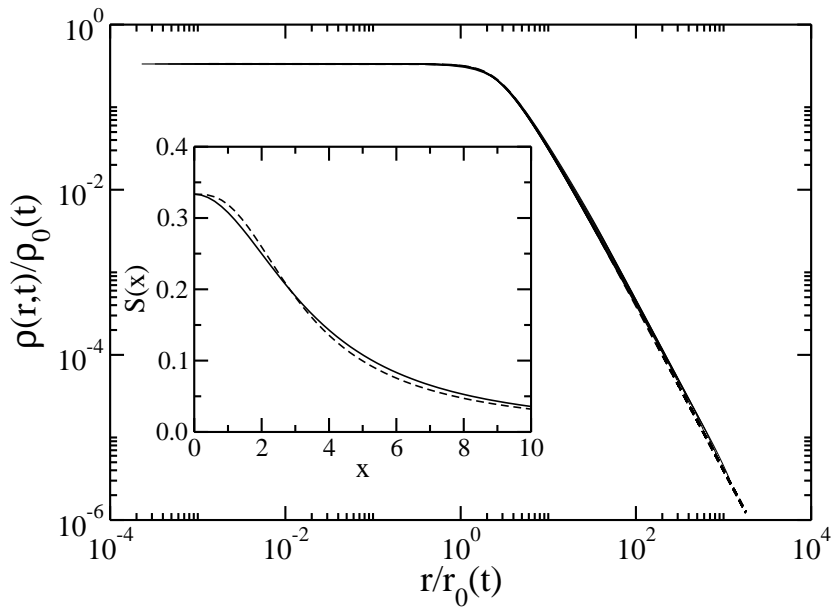}}
\caption{In the post-collapse regime, we plot $\rho(r,t)/\rho_0(t)$ as
a function of the scaling variable $x=r/r_0(t)$. A good data collapse
is obtained for central residual densities in the range $10^3\sim
10^6$. This is compared to the numerical scaling functions computed
from Eq.~(\ref{scasTn01}) (dashed line). The insert shows the
comparison between this post-collapse scaling function (dashed line)
and the scaling function below $t_{coll}$ which has been rescaled to
have the same value at $x=0$, preserving the asymptotics:
$S(x)=(3+x^2/4)^{-1}$ (see Eq.~(\ref{solscad}); full line). Note that
the post-collapse scaling function is flatter near $x=0$, as
$S(x)-1/3\sim x^3$ (in $D=3$) instead of $S(x)-1/3\sim x^2$, below
$t_{coll}$. }
\label{fig7}
\end{figure}

In the more general case $T\ne 0$, we will proceed in a very
similar way as in the previous section. We define again,
\begin{equation}
s(r,t)=\frac{M(r,t)-N_0(t)}{r^D},
\end{equation}
where $N_0$ still satisfies
\begin{equation}
\frac{d N_0}{d t}=\rho_0 N_0. \label{N0Tn0}
\end{equation}
We now obtain
\begin{equation}
{\partial s\over\partial t}=T\left(\frac{\partial^2 s}{\partial
r^2}+\frac{D+1}{r}\frac{\partial s}{\partial r}\right) +\biggl
(r{\partial s\over\partial r}+Ds\biggr )s+\frac{N_0}{r^D}\biggl
(r{\partial s\over\partial r}+Ds-\rho_0\biggr ). \label{sTn0}
\end{equation}
By definition, we have again $s(0,t)=\rho_0(t)/D$.

We look for a scaling solution of the form
\begin{equation}
s(r,t)=\rho_{0}(t)S\biggl ({r\over r_{0}(t)}\biggr ),
\label{st0post}
\end{equation}
with $S(0)=D^{-1}$. As before,  we
define the King's radius by
\begin{equation}
r_0=\left(\frac{T}{\rho_0}\right)^{1/2}.
\end{equation}
For $t<t_{coll}$, we had $s(r,t)\sim 4Tr^{-2}$ (or $S(x)\sim
4x^{-2}$). In a very short time after $t_{coll}$, this property
should be preserved, which implies that the post-collapse scaling
function should also behave as
\begin{equation}
S(x)\sim 4x^{-2},\label{condas}
\end{equation}
for large $x$. Inserting the scaling $ansatz$ in Eq. (\ref{sTn0}), we obtain
\begin{equation}
{1\over 2\rho_0^{2}}\frac{d\rho_0}{dt}\left(2 S+x S'\right)= S''
+\frac{D+1}{x}S'+S(DS+xS')+
\frac{N_0}{\rho_0r_0^D}{1\over x^{D}}(DS+xS'-1). \label{scasTn0}
\end{equation}
Again, this equation should be time independent for scaling to
hold, which implies that there exists a constant $\mu$  such
that
\begin{equation}
N_0=\mu \rho_0r_0^D.
\end{equation}
This leads to the universal behavior
\begin{equation}
\rho_0(t)=\left(\frac{D}{2}-1\right)(t-t_{coll})^{-1}.
\end{equation}
We thus end up with the scaling equation
\begin{equation}
\frac{1}{D-2}\left(2 S+x S'\right)+S'' +\frac{D+1}{x}S'+S(DS+xS')+
\mu x^{-D}(DS+xS'-1)=0,\label{scasTn01}
\end{equation}
where $\mu$ has to be chosen so that $S(x)$ satisfies the
condition of Eq.~(\ref{condas}). Its value must be determined
numerically.  Note that for small $x$, the pre-collapse scaling
function satisfies $S(x)-S(0)\sim x^2$, whereas the post-collapse
scaling function behaves as
\begin{equation}
S(x)-S(0)\sim x^D.\label{smallx}
\end{equation}
However, contrary to the $T=0$ case, $S(x)$ is not purely a
function of $x^D$.

Finally, we find that the weight of the central peak has a
universal behavior for short time after $t_{coll}$
\begin{equation}
N_0(t)=\mu\left(\frac{2}{D-2}\right)^{D/2-1}T^{D/2}\,(t-t_{coll})^{D/2-1}.
\label{n0post}
\end{equation}
Note that $N_0(t)$ behaves in a very similar manner to the mass
within a sphere of radius $r_0$ below $t_{coll}$, shown in
Eq.~(\ref{m0}). The behavior of $N_0(t)$ is illustrated in
Fig.~\ref{fig6}, while the scaling regime is displayed in
Fig.~\ref{fig7}.

In addition, comparing Eq.~(\ref{n0post}) and
Eq.~(\ref{n0postt0}), we can define again a post-collapse
cross-over time between the $T\ne 0$ and $T=0$ regimes
\begin{equation}
t_*-t_{coll}\sim T^{D/2},
\end{equation}
which is similar to the definition of Eq.~(\ref{tst0}).

\vskip 0.5cm
\begin{itemize}
\item{\it Large time limit for $T>0$}
\end{itemize}
For very large time, that is when almost all the mass has
collapsed at $r=0$, so that $N_0(t)\approx 1$, the residual
density satisfies
\begin{equation}
\rho(r,t)\sim {\rm e}^{-\lambda t}\psi(r),\label{decro}
\end{equation}
where $\psi$ satisfies the eigenequation
\begin{equation}
-\lambda \psi(r)=T \left(\psi'' +\frac{D-1}{r}\psi'\right)
+{1\over r^{D-1}}\psi'.\label{eigen1}
\end{equation}
\begin{figure}
\vskip 0.8cm
\centerline{
\psfig{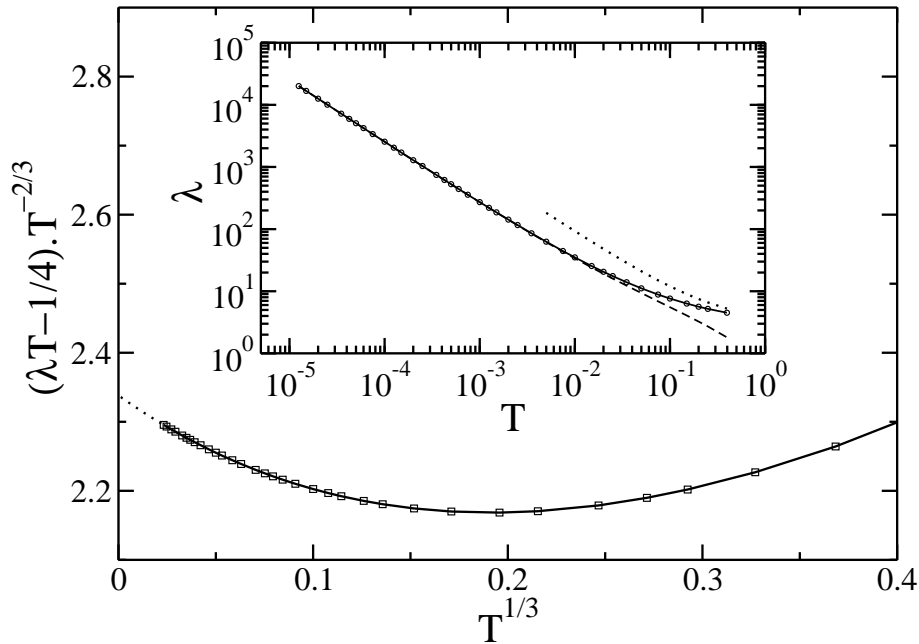}} \caption{We
plot $\lambda$ as a function of temperature (insert). The dashed
line is the small temperature expansion of Eq.~(\ref{devl}),
whereas the dotted line is the large temperature estimate which is
not very accurate in the physically relevant region $T<T_c$. The
main plot represents $\left(\lambda T-\frac{1}{4}\right) T^{-2/3}$
as a function of $T^{1/3}$ (line and squares), which should
converge to $c_{D=3}= 2.33810741...$ according to
Eq.~(\ref{devl}). We find a perfect agreement with this value
using a quadratic fit (dotted line).} \label{fig8}
\end{figure}

We did not succeed in  solving analytically the above eigenequation,
and for a given temperature, this has to be solved numerically.
However, in the limit of very small temperature, we can apply
techniques reminiscent from semiclassical analysis in quantum
mechanics ($T\leftrightarrow\hbar$). We now assume $T$ very small
and define $h(r)$ such that
\begin{equation}
\psi(r)=\exp\left[-\frac{\int_0^r h(x)\,dx}{T}\right].
\label{change}
\end{equation}
The function $h$ satisfies the following non-linear first order
differential equation
\begin{equation}
T\left(h'+\frac{D-1}{r}h\right)+\frac{h}{r^{D-1}}-h^2=\lambda
T,
\label{heq}
\end{equation}
with the simple boundary condition
\begin{equation}
h(1)=1.\label{cond}
\end{equation}
In the limit $T\to 0$, the term proportional to $T$ in the
left-hand side of Eq.~(\ref{heq}) can {\it a priori} be discarded
leading to \cite{cs3}
\begin{equation}
h(r)=\frac{2\lambda Tr^{D-1}}{1+\sqrt{1-4\lambda
Tr^{2(D-1)}}},\label{h0}
\end{equation}
which is valid for $1-r\gg T^{2/3}$. Solving perturbatively
Eq.~(\ref{heq}) leads to
\begin{equation}
\lambda=\frac{1}{4T}+\frac{c_D}{T^{1/3}}+...\qquad (T\rightarrow 0),
\label{devl}
\end{equation}
where $c_D$ is a $D$ dependent constant.

In the inverse formal limit of large temperature (although in
practice $T<T_c$), we obtain
\begin{equation}
\lambda=D+{D^{2}\over 2(D+2)}{1\over T}+...\qquad (T\rightarrow +\infty).
\label{add5c}
\end{equation}
The results of  Eq.~(\ref{devl}) and Eq.~(\ref{add5c}) are illustrated on 
Fig.~\ref{fig8}.

\section*{4. Conclusion}

$$
$$
\vskip -0.5cm

In this paper, we have illustrated the rich properties of a Brownian
model of gravitational collapse in both canonical and microcanonical
ensembles. Quantitative results have been obtained for any dimension
of space $D\ge 2$ (including the critical dimension $D=2$) and any
temperature $T\le T_{c}$ (including the peculiar case $T=0$). In the
microcanonical ensemble, we have shown that although the scaling
equation possesses solutions corresponding to $\alpha>2$, the solution
with $\alpha=2$ ultimately prevails. However, we have argued that in a
model for which the temperature is not kept uniform, we should expect
a value of $\alpha$ greater than 2 to be selected. In the canonical
ensemble, we have also shown that the singular point $t=t_{coll}$ is
not the final state of the dynamics, which is consistent with
thermodynamical considerations which predicts a totally condensed
state at $r=0$ at equilibrium \cite{kiessling,chavcano}. We have
investigated this post-collapse regime analytically, which displays
backward scaling solutions. Finally, using semiclassical methods, we
have described the very large time regime analytically. Note that the
post-collapse regime in the microcanonical ensemble is more
complex. It is marked by the formation of a central body with small
mass and small radius but with huge potential energy. This structure
is reminiscent of a ``binary star'' in astrophysics. It is surrounded
by a very hot halo with $T\rightarrow +\infty$ that is almost
homogeneous. This ``binary-halo'' structure is the most probable
structure in the microcanonical ensemble as its entropy $S\sim \ln T
\rightarrow +\infty$ \cite{cs1}. Thus, it should be reached 
in the post collapse regime. However, the Smoluchowski-Poisson system
becomes ill-defined as $T=\infty$ so that the evolution of the system
after $t_{coll}$ is pathological and requires a small-scale
regularization \cite{cs3}.  When a small-scale cut-off $h$ is
introduced, it is found that the mass of the core decreases as $h$
decreases while the temperature increases. This corroborates the
previous qualitative discussion and gives a hint as to how a rigorous
description of the post-collapse regime could be undertaken.

Except for some exact results, most of our analytical results have
been obtained by perturbative (Eq.~(\ref{devl}),
Eq.~(\ref{add5c}),...) or non perturbative methods (Eq.~(\ref{zd2}),
Eq.~(\ref{alphaexp2}),...). We hope that mathematicians will find the
following problems interesting to study in a more rigorous way.
\begin{itemize}
\item Concerning the microcanonical collapse scaling equation, it
would be interesting to justify the existence of the function
$\alpha[S(0)]$ (or $S(0)[\alpha]$), leading to a maximum value
$\alpha=\alpha_{max}$ for physical solutions.

\item The correct mathematical definition of the post-collapse stage
following the singular point $t=t_{coll}$, and the justification of
our supposedly exact estimates for $\rho_0(t)$ and $N_0(t)$ are
certainly needed.

\end{itemize}

\end{document}